\documentclass[letterpaper]{article}
\usepackage{natbib}
\usepackage{color}
\newcommand{\be}{\begin{eqnarray}}
\newcommand{\ee}{\end{eqnarray}}
\usepackage{graphicx}
\usepackage{lineno}
\modulolinenumbers[2]
\usepackage{setspace}
\title{Trade-offs drive resource specialization and the gradual establishment of ecotypes}

\author{Bj{\o}rn {\O}stman$^{1,2,*}$, Randall Lin$^{3}$, \and Christoph Adami$^{1,2}$\\
\mbox{}\\
$^1$Department of Microbiology and Molecular Genetics\\
Michigan State University, East Lansing, MI 48824\\
$^2$BEACON Center for the Study of Evolution in Action\\
Michigan State University, East Lansing, MI 48824\\ 
$^3$California Institute of Technology, Pasadena, CA 91125\\
ostman@msu.edu, rllin@caltech.edu, adami@msu.edu\\
\\
$^*$ Corresponding author
adaptive radiation; gradualism; phyletic evolution \\
}
 
\begin{document}
\maketitle

\newpage

\begin{abstract}\textbf{Background}\\
Speciation is driven by many different factors. Among those are trade-offs between different ways an organism utilizes resources, and these trade-offs can constrain the manner in which selection can optimize traits. Limited migration among allopatric populations and species interactions can also drive speciation, but here we ask if trade-offs alone are sufficient to drive speciation in the absence of other factors.\\

\textbf{Results}\\
We present a model to study the effects of trade-offs on specialization and adaptive radiation in asexual organisms based solely on competition for limiting resources, where trade-offs are stronger the greater an organism's ability to utilize resources. In this model resources are perfectly substitutable, and fitness is derived from the consumption of these resources. The model contains no spatial parameters, and is therefore strictly sympatric. We quantify the degree of specialization by the number of ecotypes formed and the niche breadth of the population, and observe that these are sensitive to resource influx and trade-offs. Resource influx has a strong effect on the degree of specialization, with a clear transition between minimal diversification at high influx and multiple species evolving at low resource influx. At low resource influx the degree of specialization further depends on the strength of the trade-offs, with more ecotypes evolving the stronger trade-offs are. The specialized organisms persist through negative frequency-dependent selection. In addition, by analyzing one of the evolutionary radiations in greater detail we demonstrate that a single mutation alone is not enough to establish a new ecotype, even though phylogenetic reconstruction identifies that mutation as the branching point. Instead, it takes a series of additional mutations to ensure the stable coexistence of the new ecotype in the background of the existing ones, reminiscent of a recent observation in the \emph{E. coli} long-term experiment.\\

\textbf{Conclusion}\\
Trade-offs are sufficient to drive the evolution of specialization in sympatric asexual populations. Without trade-offs to restrain traits, generalists evolve and diversity decreases. The observation that several mutations are required to complete speciation, even when a single mutation creates the new species, highlights the gradual nature of speciation and the importance of phyletic evolution.
\end{abstract}

\newpage

\section*{Introduction}
Trade-offs present limitations to the adaptive potential of organisms and are commonly thought of as the reason why we observe such an abundance of species, rather than just a few that have adapted to any conditions that life may present~\citep{agrawaletal2010}. While the divergence of allopatric species is sustained by geographic barriers, sympatric populations can split through the action of negative frequency-dependent selection. This force for diversification can be driven by fitness trade-offs between different niches, without which a single generalist phenotype could sweep the population.
	Empirically two- and three-way trade-offs have been observed in several microbial systems, such as rate vs. yield in \emph{E. coli}~\citep{Novaketal2006}, nitrogen and phosphorous affinity and cell volume in phytoplankton~\citep{Edwardsetal2011}, or ability to adapt to a CO$_2$ enriched environment and competitive ability~\citep{Collins2011}. Modeling approaches have explored the trade-offs between two traits, such as rate of resource acquisition vs. biomass production (rate vs. yield) and maximum uptake rate vs. affinity~\citep{Frank2010,Gudeljetal2007}, leading to co-existence of two ecotypes on one resource.

To address the influence of trade-offs and other factors on diversification in asexual organisms, we study a model where fitness gained from resources is dependent on explicitly modeled trade-offs between the traits that control resource use. We aim to quantify the impact that trade-offs have on the degree of resource specialization, measured as the number of distinct ecotypes that can co-exist. Fitness is given as a function of the organism's ability to utilize the available resources modeled on the Monod equation~\citep{Monod1949}, modified such that having high affinity for more resources comes at a cost in fitness leading to trade-offs. Spatial structure is absent from the model, making the system strictly sympatric, as opposed to weak sympatry, which retains a spatial component that can affect the dynamics of the system (e.g.,~\citealp{KoppHermisson2008,GavriletsWaxman2002,KondrashovKondrashov1999}).

We restrict our model to asexual organisms, showing how resource specialization leads to adaptive radiation in the absence of reproductive isolation. This best mirrors evolutionary and ecological dynamics of unicellular, asexual organisms in which mating and genetic recombination are absent. Such organisms include bacteria, in which adaptive radiation has been observed on several occasions~\citep{MacLeanetal2005,RaineyTravisano1998,Halletal2007}, and plankton, which forms the basis for the question of how several species occupying the same niche can coexist seemingly indefinitely, the so-called \emph{paradox of the plankton}~\citep{Hutchinson1961,Kolesov1992,HornCattron2003,Petersen1975,RoyChattopadhyay2007,Schippersetal1974,Shoresh2008, Sommer1984}.

We determine the number of species using the \emph{Ecological Species Concept}~\citep{VanValen1976, Konstantinidisetal2006, Wilkins2007}, which defines a species as ``a lineage which occupies an adaptive zone minimally different from that of any other lineage in its range and which evolves separately from all other lineages outside its range," without claiming that this definition is always appropriate. Species by this definition are also denoted \emph{ecotypes}, and here we use the two terms interchangeably.

Depending on the parameters governing resource abundance, mutational effects, and trade-offs leading to negative frequency-dependent selection, we observe resource competition giving rise to either generalists or specialists. By tracking the evolving organisms and reconstructing the phylogenetic relationship of the surviving lineages we can identify the exact mutations that cause the initial divergence between lineages. This analysis then enables us to distinguish between anagenetic and cladogenetic change, addressing a long-standing question about whether macroevolutionary change proceeds by gradualistic or punctuated modes (e.g.~\citealp{Pennelletal2014}).

\section*{Results}
\begin{figure}[htp]
\begin{center}
\includegraphics[width=3.3in,angle=0]{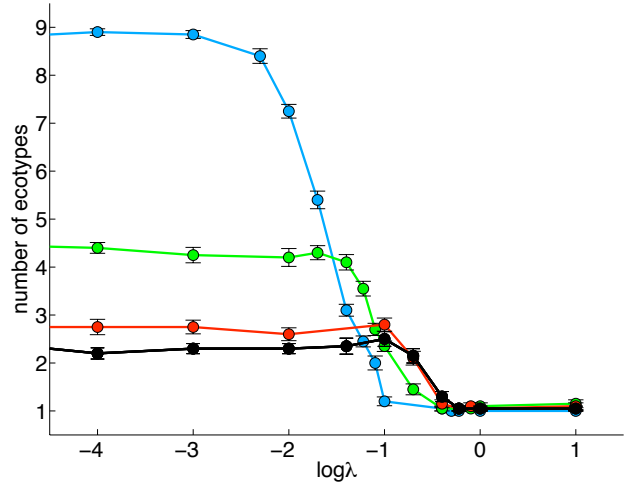}
\caption{Number of ecotypes as a function of resource influx for different levels of trade-off. A low influx creates a pressure for the population to split into different ecotypes
because it this enables selection to differentiate between phenotypes that use different resources at different abundances. The degree of specialization does not diminish when the influx decreases (see Discussion). More specialists evolve when trade-offs are more severe, essentially creating one niche per resource for the highest cost. Even when having a high resource affinity is not penalized, the population still fragments into on average two stable lineages.
Blue markers: ($\sigma_1$,$\sigma_2$) = (10,1), green: ($\sigma_1$,$\sigma_2$) = (1,0.1), red: ($\sigma_1$,$\sigma_2$) = (0.1,0.01), black: ($\sigma_1$,$\sigma_2$) = (0,0). $N=1000$, $\mu=0.05$. Each datum is the average of 20 simulations, and error bars are s.e.m.}
\label{fig_nt}
\end{center}
\end{figure}

We carried out simulations with different values of resource influx, $\lambda$, cost-parameters, $\sigma_1$ and $\sigma_2$, population size, $N$, and mutation rate, $\mu$ and examined their impact on the degree of specialization. By far the two biggest factors affecting the degree of specialization are the resource influx and trade-offs (Fig.~\ref{fig_nt}). The number of ecotypes $n_t$ is first and foremost a function of the amount of resources that flow into the system. When this rate is high, the number of ecotypes is close to one, and when it is low, the number of types increases. Contrary to results from experiments with bacteria~\citep{Halletal2007} and digital evolution~\citep{Chowetal2004}, which both found unimodal distributions of diversity as a function of resource influx, here the number of types does not decrease even for very low influx (see Discussion). When the influx is low, the number of ecotypes is strongly dependent on the cost-parameters, with high cost leading to more ecotypes. As the influx increases there is a transition where the number of ecotypes drops to one. When the resource influx is high, generalists dominate, and adaptive radiation and specialization do not occur. The dynamics of the model depends on the resource abundance, such that selection can only differentiate between phenotypes using different resources when those resource differences result in fitness differences that are larger than the inverse of the population size, $s \geq 1/N$ (e.g.,~\citealp{Gillespie2004}). If the fitness difference between two organisms is smaller than this value, then genetic drift has a larger influence on the dynamics, and selection will not favor either organism. In that case, an incentive for the population to split into different groups ceases to exist, and generalists evolve. The more severe the cost is, the more specialists evolve, essentially creating one niche per resource for the highest cost. Even when there is no cost of having high resource affinity, the population still fragments into approximately two stable lineages, but this is an effect of fixation by drift being a very slow process. When the zero-mutation rate experiments are run for a much longer time, the most dominant phenotype eventually replaces all other leaving just one ecotype.


We estimated the \emph{most recent common ancestor} (MRCA) by tracing the lines of descent backwards and noting when they coalesce. If a population does not split into more than one stable subpopulation, the MRCA should be relatively recent. When lineages coexist for a long time, the MRCA will be far in the past. In most cases the first split occurs very early in the simualtions, i.e., two or more organisms in the starting population start lineages that persist until the end of the runs. We also ran simulations in a neutral fitness landscape, where all organisms have the same fitness, and found that the MRCA is always very close to the present. In this case genetic drift causes the population to descend from a single ancestor not very far in the past, as expected.

The degree of specialization does not change when the simulation is run longer. When the simulations are run for $80,000$ updates (see Methods) before turning off mutations, neither $n_t$ nor $B$ changes significantly compared to running them for $40,000$ updates. What does change in between these two lengths of the simulations is the mean resource affinity, which continues to increase, except in simulations in which high affinities are heavily penalized (e.g., $\sigma_1 = 10$).

Trait-lethals have a significant impact on the degree of specialization. Without the possibility of traits mutating to zero, niche breadth is nearly always at the maximum of one, and $n_t$ is severely depressed compared to simulations that have trait-lethal mutations (table~\ref{table_nolethals}). For the small population size of $N=100$ the number of ecotypes was consistently much lower than for $N=1000$. In the absence of trait-lethal mutations none of the parameter sets tried gave an average $n_t$ greater than $2$.  Increasing the population size to $N=5000$ results in only slightly more ecotypes compared to $N=1000$ (table~\ref{table_N}). Trait-lethal mutations make it difficult to increase the affinity of all traits at the same time, because they reduce the affinities to zero. Phenotypes that have high affinity for many resources therefore become rare in the population. However, among the organisms on the line of descent very few instances of trait-lethals that decrease affinity by more than 0.1 are observed (the most extreme case observed was from 0.2 to 0).

\begin{figure}[htp]
\begin{center}
\includegraphics[width=3.3in,angle=0]{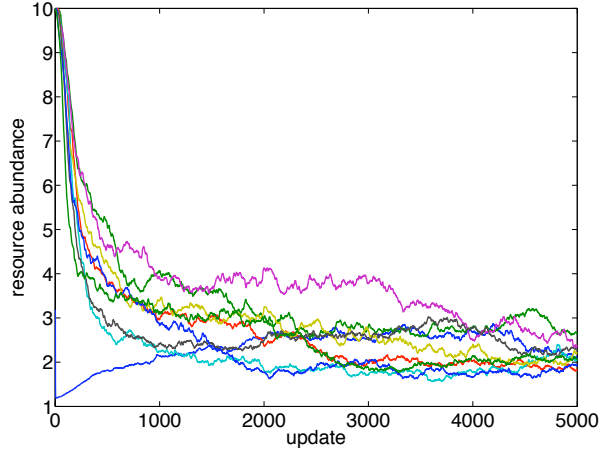}
\caption{Resource abundance for nine resources as a function of time. $N=1000$, $\mu=0.05$, $\lambda=10^{-3}$, $\sigma_1=1$, $\sigma_2=0.1$.}
\label{fig_resources}
\end{center}
\end{figure}

Simulations start with a homogeneous population consisting of specialists for one resource ($c_1=0.1$). However, starting with generalists that have a non-zero affinity for all resources results in the same degree of specialization (table~\ref{table_generalists}). Resource abundances usually equilibrate within a few hundred updates (Fig.~\ref{fig_resources}), and hence the number of ecotypes is unaffected by the resource abundance at the beginning of the simulations.\\


\subsection*{Gradual establishment of ecotypes}
\begin{figure}[htp]
\begin{center}
\includegraphics[width=3.3in,angle=0]{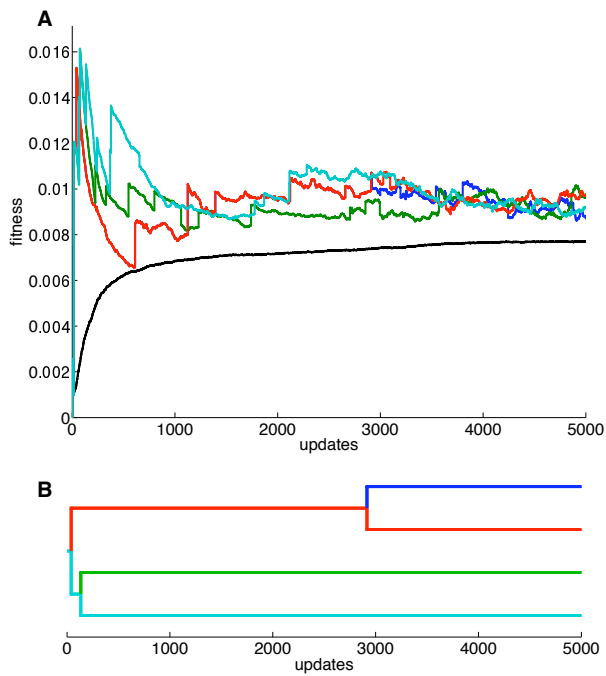}
\caption{Simulation resulting in the four ecotypes shown in table~\ref{table_four_phenotypes}. (A) Fitness as a function of time for each of the four ecotypes (blue, red, green, cyan) and the population mean fitness (black). (B) Phylogram showing the evolutionary relationship between the four ecotypes. $N=1000$, $\mu=0.05$, $\lambda=10^{-3}$, $\sigma_1=1$, $\sigma_2=0.1$, starting with a homogeneous population of specialists with affinity equal to 0.1 for the first resource and zero for the rest.}
\label{fig_four_LODs}
\end{center}
\end{figure}

In order to gain insights into the mechanics of specialization, we reconstructed all mutations on the line of descent for a simulation that resulted in four ecotypes. This enabled us to track the changes that eventually lead to diversification, record how early they arise, and whether they appear gradually or within a short time-span. Because tracking all $N$ phenotypes is computationally demanding, we were not able to run simulations longer than 5,000 updates when reconstructing the lines of descent. However, the splitting into distinct stable phenotypes always happens much earlier than this, and the shorter run therefore does not impact these findings. In Fig.~\ref{fig_four_LODs} we show the lines of descent in a simulation that after 5,000 updates resulted in four stable phenotypes (table~\ref{table_four_phenotypes}).

Tracking these four ecotypes revealed that the first two bifurcations happened very early on, while the last took place at update 2,916 (table~\ref{table_four_lod}). The first mutational event is a double mutation that increases affinity for the seventh and ninth resources at update 15. This is an extremely advantageous mutation, with a selection coefficient of $s=11.271$. Not surprisingly, this organism swept to fixation, leading to fast depletion of those resources. As a consequence those resources conferred considerably less fitness than when the organism was rare. The split between the red and cyan lineages in Fig.~\ref{fig_four_LODs} was first marked by an increase in affinity in the red lineage for resource three at update 40. However, at this point the red lineage could have outcompeted the cyan lineage, but an increase in affinity for resource two at update 67 ensured that the cyan lineage had an advantage over the red lineage in utilizing this resource. Only with both of these mutations can the two phenotypes coexist through negative frequency-dependent selection. Similarly, the split between the green and cyan lineages at update 134 does not result in two separate ecotypes until at least one mutation later in the green lineage at update 225. The split between the red and the blue lineages also required at least two mutations at updates 2,916 and 2,999 to establish these two diverging phenotypes as ecotypes. This mirrors the gradual emergence of a stable polymorphism in \emph{E. coli}, where three separate mutations in regulatory genes were needed to produce frequency-dependent effects~\citep{Plucainetal2014}.


\subsection*{Variable population size}
To examine the effects of assuming a constant population size we ran the simulations with a variable population size as a control. In this instance of the model, all individuals are given a chance to reproduce equal the their fitness. Because fitness given by eq.~(\ref{eq_fitness}) is always between zero and one, we can use fitness as this probability. At low resource influx, the population occasionally goes extinct when it cannot diversify from the initial specialist phenotype fast enough to gain high fitness. We therefore started the variable population size simulations with a homogeneous population of generalists where all affinities are equal to $0.1$. Note that whether one starts with generalists or specialists makes no difference for the degree of specialization (table~\ref{table_generalists}). 

The variable population size simulation is very sensitive to the balance between the resource influx and decay. If the decay is too high or the influx too low, the population quickly goes extinct. On the other hand, if the influx is very high, or decay is low, the population quickly grows to sizes beyond $3,000$, which makes the simulation unwieldy. For these reasons we were unable to run the variable population size simulations with resource influx lower than $\lambda=2\cdot10^{-4}$ and higher than $4\cdot10^{-3}$. Within this range the population becomes stable when the rate at which individuals are removed ($1\%$) is equal to the average fitness. This range of influx values results in population sizes ranging from $N=100$ to $2,520$ (table~\ref{table_varpop}).

Results for variable population sizes are comparable to those with a constant population size. A larger population size does increase the number of ecotypes (table~\ref{table_N}), but there is no significant difference between the constant and variable population size simulations as long as the stable population size in the variable treatment is the same as the constant population size. 

\section*{Discussion}
Trade-offs are ubiquitous in nature as species wrestle with the benefits and drawbacks of trait value optimization. In the absence of trade-offs the populations would evolve to become a generalist ``superspecies"~\citep{Tilman1982}, and ultimately a few species would dominate, with extinction, geography and stochastic processes being the only motors of diversity. Antagonistic pleiotropy was previously found to be the primary cause of resource specialization and niche breadth reduction in Avida~\citep{Ostrowskietal2007}, but trade-offs in Avida cannot be modified to investigate this effect quantitatively. We therefore studied sympatric, asexual populations using an individual-based model with discrete traits where fitness is an explicit function of resource consumption, and added trade-offs to this model by adding a simple of a cost-function to the Monod equation. We found that populations of asexual organisms in sympatry fragment into specialist ecotypes via adaptive radiation, with the degree of specialization determined largely by the severity of trade-offs. Diversification consistently occurs when resources vary enough that selection can distinguish between different phenotypes, and negative frequency-dependent selection can prevent rare phenotypes from being outcompeted. The action of negative frequency-dependent selection is contingent on the presence of trade-offs to give specialists a fitness advantage over generalists.

The origin and maintenance of generalists is only observed at high influx, while specialists dominate at low influx. High resource influx results in a population of generalists because in this case there is little competition for resources. In that case the resource abundance is always so high that both the resource maximum and minimum, $N\lambda/c_0$ and $N\lambda$, respectively, give $r_k/(r_k+\gamma_k)$ close to one (eq.~\ref{eq_fitness}). When this happens, the amount of resource available makes no difference for selection, and there is thus no benefit to losing affinity in order to reduce cost, because all resources are abundant enough that using them at all is advantageous. On the other hand, when the resource influx is low, then the resource abundance is always low compared to the half-saturation, $\gamma_k$, and any change in resource abundance affects fitness nearly linearly. Thus, at low resource influx selection can easily differentiate between an organism that only uses a scarce resource from one that uses an abundant one. We can also state this by saying that specialization does not happen when the population is always well-fed, but it occurs readily when the population is on the brink of starvation. Only when resource abundance is generally low does the environment induce the population to diversify.

The sustained degree of specialization at low influx is unlike experiments in the digital evolution platform, Avida~\citep{Chowetal2004}, where specialization was observed only at intermediate resource influx, and not at very low or very high influx. In Avida the reduction in degree of specialization at low influx is likely due to the fact that organisms can reproduce at a low rate even in the absence of resources. This makes it difficult for selection to distinguish between different phenotypes, and thus impairs the action of negative frequency-dependent selection to sustain specialization. In our study the effect of having a constant population size is that only relative fitness matters, and relative fitness is only minimally affected by lowering the influx. A model with a forced constant population size therefore does not capture the fact that low absolute fitness, caused by fewer available resources, should decrease the total reproductive output of the population, and thereby decrease the population size. When we relaxed this assumption the population became very sensitive to the exact rate of resource flowing into the system. Indeed, we found that the range of influx that can support a population is quite narrow. Low resource influx quickly leads to extinction, while high influx lead to a population explosion that quickly becomes difficult to handle computationally. For small population sizes selection is unable to differentiate between individuals with different phenotypes. As a consequence the population drifts, disabling negative frequency-dependent selection, which is otherwise the motor of specialization. Apart from this difference, when the influx gives rise to stable populations, the variable population size implementation gives results comparable to those simulation in which the population size is fixed.

The route by which adaptive radiation occurs is very informative about the evolution of specialization. By reconstructing the complete evolutionary history of each of the surviving ecotypes in a single simulation, we can track the exact mutational changes on the line of descent. The changes in phenotype over time show that when one ecotype splits into two (i.e., at the moment of incipient speciation), it does so by mutational changes to the phenotype that in hindsight are not enough to sustain the split. If the first change that separates two lineages would be the only difference between them, one of them would always outcompete the other in a zero-mutation rate experiment. Only through continued phyletic evolution will the first mutations lead to specialization and an increase in the number of ecotypes.

Trait-lethal mutations were introduced because it is generally easier to destroy function than it is to create it. If a pathway for a particular function involves a hundred genes, it may only take a single mutation to disrupt or silence an essential gene, rendering the whole pathway non-functional (note that genetic robustness will dampen this effect~\citep{Kitano2004}). A mutational scheme with trait-lethality, where 70\% of mutations cause loss of function, biases the effect of mutations towards specialists. Compared to the simulation results without these trait-lethal mutations, the degree of specialization is significantly lower when one mutation cannot destroy the function. This suggests that specialization and niche breadth reduction are amplified by this mechanism, whether it be through destruction of weakly robust pathways and modules (as simulated here), or through antagonistic pleiotropy, as observed in Avida~\citep{Ostrowskietal2007}.

\subsection*{Conclusion}
In the model presented here it is clear that trade-offs are needed for specialization and adaptive radiation to occur. Sex and spatial heterogeneity can drive specialization and diversification (e.g.,~\citealp{Weissingetal2011,deAguiaretal2009}), but here we see that they are not necessary components as long as trade-offs in resource utilization are present. Specialization happens when resources are scarce, but only when the benefit of utilizing resources is constrained by trade-offs.

We have outlined an example of gradual, phyletic evolution wherein the first steps toward speciation do not in themselves complete speciation. Different lineages can only be sustained as ecotypes by negative frequency-dependent selection after continued specialization on different resources. Mechanistic insight into incipient speciation could be gained by quantifying the effects of zero-mutation rate experiments shortly after such events occur.

Trade-offs are often created on a genetic level by antagonistic pleiotropy. Since there is no genetics in the Monod model, we have instead modeled trade-offs by an explicit mathematical function. Trade-offs could be modeled within a framework that has an explicit genetic basis that includes epistasis and pleiotropy, such as the NK model~\citep{Ostmanetal2012}. In such a model genetic constraints can emerge naturally from the interaction between genetic elements, much as in the emergence of Dobzhansky-Muller incompatibilities~\citep{Orretal2001}, which will make it possible to study a wider range of genotype-phenoptype map effects on speciation and the creation of biological diversity.

\section*{Methods}
We simulate evolution by subjecting organisms to reproduction, mutation, and selection. Each organism $i$ has a phenotype consisting of as many traits as there are resources, $R$. Each trait is a \emph{resource affinity} $c_{ik}$, which describes the efficiency with which an organism utilizes resource $k$. The population size $N$ is kept constant by randomly removing $1\%$ of the population every computational update (equivalent to about 100 updates per generation), and replacing those organisms by randomly choosing which survivors reproduce with a probability based on their relative fitness. This replacement scheme is equivalent to a death-birth Moran process with multiple deaths/births per update (see e.g.,~\citealp{Ostmanetal2012,AdamiHintze2013}). Offspring inherit the phenotype of the parent, but every trait value $c_{ik}$ has a chance $\mu$ of mutating, resulting either in a decrease or an increase in resource affinity. Resource affinities vary between 0 (minimum utilization) and $1$ (maximum utilization). $30\%$ of mutations change a trait value by either increasing or decreasing resource affinity by $0.1$ with equal probability. $70\%$ of mutations are \emph{trait-lethals}, which have the effect of setting the resource affinity to zero. This is meant to stimulate diversification by disrupting the function of utilizing one distinct resource. Once a trait is affected by a trait-lethal mutation, a subsequent mutation may again increase its value by $0.1$. 

Resource competition among organisms is modeled in a way similar to the standard model based on the Monod equation ~\citep{Monod1949,Shoresh2008, HuismanWeissing1999}, which describes fitness of each phenotype in the model. Individual fitness is a function of affinity, $c_{ik}$, to each resource, $k$, the resource abundance, $r_k$, and of how costly it is to have a non-zero affinity (eq.~\ref{eq_fitness}). The half-saturation, $\gamma_k$, is the resource abundance at which fitness from resource $k$ is half of its maximal value:
\be
F_i = \frac{1}{1+D_i} \sum_{k} \frac{c_{ik}r_k}{(r_k+\gamma_{k})}. 	\label{eq_fitness}
\ee
The term $1/(1+D_i)$ is a sum over resource affinities that simulates trade-offs by assigning a cost to positive ($\sigma_1$) and in particular multiple ($\sigma_2$) resource affinities:
\be
D_i = \sum_k (\sigma_1 c_{ik} + \sigma_2\xi(c_{ik})).	 \label{eq_cost}
\ee
Here $\sigma_1$ specifies the cost of having a high affinity, while $\sigma_2$ specifies the cost of using many resources without differentiating between high and low affinity. $\xi(c_{ik})$ is 0 for $c_{ik}=0$ and $1$ otherwise, therefore this function does not distinguish between high and low affinity, but only between zero and non-zero affinity.

As in Michaelis-Menten kinetics, each resource is replenished every update at a constant influx per individual $\lambda_k$, decays at a rate $c_0$ proportional to the amount of available resource, and is consumed by the organisms at a rate equal to the total fitness in the population derived from that resource, $\sum_iF_{ik}$, where $F_{ik}$ is the fitness derived from the $k$th resource by the $i$th individual (eq.~\ref{eq_resources}).
\be
\frac{dr_k}{dt}=N\lambda_k-c_0r_k-\sum_iF_{ik}.		 \label{eq_resources}
\ee
Resource abundance is at a maximum when there is no consumption, in which case we solve eq.~(\ref{eq_resources}) for $\frac{dr_k}{dt}=0$ with $\sum_iF_{ik}=0$ and obtain $r=N\lambda/c_0$. When a resource is fully consumed every update before it can decay, the resource abundance is just equal to the amount of resources that flows into the system. When the decay term is zero, the resource abundance is given by the influx, $r=N\lambda$. 

\subsection*{Degree of specialization}
With a mutation rate of $\mu=0.05$, phenotype space is fairly well explored, consistently resulting in hundreds of unique phenotypes (i.e., unique combinations of resource affinities) within a few hundred updates. To find the number of \emph{ecotypes} in the population after $40,000$ updates of reproduction, we identify stable phenotypes by performing a \emph{zero-mutation rate experiment}. With the mutation rate set to to zero no new phenotypes are generated, and existing phenotypes begin to outcompete each other. We continue simulations until the remaining phenotypes cannot outcompete each other (no more than $80,000$ updates). If this steady-state of the population was in doubt, we performed \emph{invasion experiments} to establish the stability of the population. Invasion experiments were performed by setting one of the phenotypes to a low frequency and checking if this phenotype had the ability to invade when rare. No inconsistencies were found between the two methods.

We quantify the \emph{degree of specialization} by the number of ecotypes $n_t$ and the population \emph{niche breadth} (e.g.,~\citealp{Devictoretal2010}), calculated as the fraction of non-zero resource affinities in the population:
\be
B = \sum_{i,k} \xi(c_{ik})/NR.
\ee
$B$ equals $1/R$ when every individual utilizes exactly one resource, and $1$ when every individual uses all resources. A population of generalists is thus characterized by a high niche breadth, while resource specialization results in low niche breadth. This definition is close to that used in investigations of niche breadth reduction in Avida~\citep{Ostrowskietal2007}, where it is calculated as the proportion of organisms performing each function summed over all functions, rather than taking the mean over all resources, as we do here.\\

\subsection*{Initial conditions}
We start simulations with a homogeneous population comprised of specialists that have one non-zero affinity $c_{i1}=0.1$ and the rest equal to zero, or with generalists that have affinity $0.1$ for all resources. Throughout we use $R=9$ resources, $c_0=0.1$, and $\gamma_k=100$ for all resources. We start with all resource abundances either at a minimum equal to the influx, $N\lambda$, or at the maximum given by eq.~(\ref{eq_resources}) without consumption, $N\lambda/c_0$, but we emphasize that distinction makes no difference to the evolution of the population, because resource abundances adjust based on consumption before the population has enough time to change how they use resources.

\section*{Competing interests}
The authors declare that they have no competing interests.

\section*{Authors' contributions}
B{\O} and RL wrote the code and carried out the simulations. B{\O} performed the data analysis. B{\O} and CA conceived of the study and participated in its design and coordination and helped to draft the manuscript. All authors read and approved the final manuscript.

\section*{Acknowledgements}
The authors wish to thank Annat Haber for extensive comments on the manuscript. This work was supported by the National Science Foundation's Frontiers in Integrative Biological Research grant FIBR-0527023 and the National Science FoundationÕs BEACON Center for the Study of Evolution in Action under contract no. DBI-0939454. Randall Lin was supported by the National Science Foundation's Research Experiences for Undergraduates grant to Keck Graduate Institute. This work was supported in part by Michigan State University through computational resources provided by the Institute for Cyber-Enabled Research.

\footnotesize
\bibliographystyle{amnatnat}
\bibliography{ecotypes}

\begin{thebibliography}{38}
\providecommand{\natexlab}[1]{#1}

\bibitem[{Adami and Hintze(2013)}]{AdamiHintze2013}
Adami, C., and A.~Hintze. 2013.
\newblock {Evolutionary instability of zero-determinant strategies demonstrates
  that winning is not everything}.
\newblock Nature Communications 4.

\bibitem[{Agrawal et~al.(2010)Agrawal, Conner, and Rasmann}]{agrawaletal2010}
Agrawal, A.~A., J.~K. Conner, and S.~Rasmann. 2010.
\newblock {Tradeoffs and Negative Correlations in Evolutionary Ecology}.
\newblock Pages 243--268 \emph{in} M.~Bell, W.~Eanes, D.~Futuyma, and
  J.~Levinton, eds. Evolution After Darwin: the First 150 Years. Sinauer
  Associates.

\bibitem[{Chow et~al.(2004)Chow, Wilke, Ofria, Lenski, and
  Adami}]{Chowetal2004}
Chow, S., C.~Wilke, C.~Ofria, R.~Lenski, and C.~Adami. 2004.
\newblock Adaptive radiation from resource competition in digital organisms.
\newblock Science 305:84--86.

\bibitem[{Collins(2011)}]{Collins2011}
Collins, S. 2011.
\newblock {Competition limits adaptation and productivity in a photosynthetic
  alga at elevated CO2}.
\newblock Proceedings of the Royal Society B: Biological Sciences 278:247--255.

\bibitem[{de~Aguiar et~al.(2009)de~Aguiar, Baranger, Baptestini, Kaufman, and
  Bar-Yam}]{deAguiaretal2009}
de~Aguiar, M. A.~M., M.~Baranger, E.~M. Baptestini, L.~Kaufman, and Y.~Bar-Yam.
  2009.
\newblock Global patterns of speciation and diversity.
\newblock Nature 460:384--387.

\bibitem[{Devictor et~al.(2010)Devictor, Clavel, Julliard, Lavergne, Mouillot,
  Thuiller, Venail, Villeger, and Mouquet}]{Devictoretal2010}
Devictor, V., J.~Clavel, R.~Julliard, S.~Lavergne, D.~Mouillot, W.~Thuiller,
  P.~Venail, S.~Villeger, and N.~Mouquet. 2010.
\newblock Defining and measuring ecological specialization.
\newblock Journal of Applied Ecology 47:15--25.

\bibitem[{Edwards et~al.(2011)Edwards, Klausmeier, and
  Litchman}]{Edwardsetal2011}
Edwards, K.~F., C.~A. Klausmeier, and E.~Litchman. 2011.
\newblock {Evidence for a three-way trade-off between nitrogen and phosphorus
  competitive abilities and cell size in phytoplankton}.
\newblock Ecology 92:2085--2095.

\bibitem[{Frank(2010)}]{Frank2010}
Frank, S.~A. 2010.
\newblock {The trade-off between rate and yield in the design of microbial
  metabolism}.
\newblock Journal of Evolutionary Biology 23:609--613.

\bibitem[{Gavrilets and Waxman(2002)}]{GavriletsWaxman2002}
Gavrilets, S., and D.~Waxman. 2002.
\newblock Sympatric speciation by sexual conflict.
\newblock Proceedings of the National Academy of Sciences of the United States
  of America 99:10533--10538.

\bibitem[{Gillespie(2004)}]{Gillespie2004}
Gillespie, J. 2004.
\newblock {Population Genetics: A Concise Guide, Second Edition}.
\newblock Johns Hopkins University Press.

\bibitem[{Gudelj et~al.(2007)Gudelj, Beardmore, Arkin, and
  MacLean}]{Gudeljetal2007}
Gudelj, I., R.~Beardmore, S.~S. Arkin, and R.~C. MacLean. 2007.
\newblock {Constraints on microbial metabolism drive evolutionary
  diversification in homogeneous environments}.
\newblock Journal of Evolutionary Biology 20:1882--1889.

\bibitem[{Hall and Colegrave(2007)}]{Halletal2007}
Hall, A.~R., and N.~Colegrave. 2007.
\newblock {How does resource supply affect evolutionary diversification?}
\newblock Proceedings of the Royal Society B: Biological Sciences 274:73--78.

\bibitem[{Horn and Cattron(2003)}]{HornCattron2003}
Horn, J., and J.~Cattron. 2003.
\newblock The paradox of the plankton: Oscillations and chaos in multispecies
  evolution.
\newblock Genetic and evolutionary computation - GECCO 2003, Pt I, proceedings
  2723:298--309.
\newblock Lecture notes in computer science.

\bibitem[{Huisman and Weissing(1999)}]{HuismanWeissing1999}
Huisman, J., and F.~Weissing. 1999.
\newblock Biodiversity of plankton by species oscillations and chaos.
\newblock Nature 402:407--410.

\bibitem[{Hutchinson(1961)}]{Hutchinson1961}
Hutchinson, G. 1961.
\newblock The paradox of the plankton.
\newblock The American Naturalist 95:137--145.

\bibitem[{Kitano(2004)}]{Kitano2004}
Kitano, H. 2004.
\newblock {Biological robustness}.
\newblock Nature Reviews Genetics 5:826--837.

\bibitem[{Kolesov(1992)}]{Kolesov1992}
Kolesov, Y. 1992.
\newblock The paradox of plankton is explained.
\newblock Biofizika 37:1113--1114.

\bibitem[{Kondrashov and Kondrashov(1999)}]{KondrashovKondrashov1999}
Kondrashov, A.~S., and F.~A. Kondrashov. 1999.
\newblock Interactions among quantitative traits in the course of sympatric
  speciation.
\newblock Nature 400:351--354.

\bibitem[{Konstantinidis et~al.(2006)Konstantinidis, Ramette, and
  Tiedje}]{Konstantinidisetal2006}
Konstantinidis, K.~T., A.~Ramette, and J.~M. Tiedje. 2006.
\newblock The bacterial species definition in the genomic era.
\newblock Philosophical Transactions of the Royal Society B 361:1929--1940.

\bibitem[{Kopp and Hermisson(2008)}]{KoppHermisson2008}
Kopp, M., and J.~Hermisson. 2008.
\newblock Competitive speciation and costs of choosiness.
\newblock Journal of Evolutionary Biology 21:1005--1023.

\bibitem[{MacLean et~al.(2005)MacLean, Dickson, and Bell}]{MacLeanetal2005}
MacLean, R.~C., A.~Dickson, and G.~Bell. 2005.
\newblock Resource competition and adaptive radiation in a microbial microcosm.
\newblock Ecology Letters 8:38--46.

\bibitem[{Monod(1949)}]{Monod1949}
Monod, J. 1949.
\newblock The growth of bacterial cultures.
\newblock Annual Review of Microbiology 3:371--394.

\bibitem[{Novak et~al.(2006)Novak, Pfeiffer, Lenski, Sauer, and
  Bonhoeffer}]{Novaketal2006}
Novak, M., T.~Pfeiffer, R.~E. Lenski, U.~Sauer, and S.~Bonhoeffer. 2006.
\newblock {Experimental Tests for an Evolutionary Trade‐Off between Growth
  Rate and Yield in E. coli}.
\newblock The American Naturalist 168:242--251.

\bibitem[{Orr and Turelli(2001)}]{Orretal2001}
Orr, H.~A., and M.~Turelli. 2001.
\newblock {The Evolution of Postzygotic Isolation: Accumulating
  Dobzhansky-Muller Incompatibilities} 55:1085--1094.

\bibitem[{{\O}stman et~al.(2012){\O}stman, Hintze, and Adami}]{Ostmanetal2012}
{\O}stman, B., A.~Hintze, and C.~Adami. 2012.
\newblock {Impact of epistasis and pleiotropy on evolutionary adaptation}.
\newblock Proceedings of the Royal Society B: Biological Sciences 279:247--256.

\bibitem[{Ostrowski et~al.(2007)Ostrowski, Ofria, and
  Lenski}]{Ostrowskietal2007}
Ostrowski, E., C.~Ofria, and R.~Lenski. 2007.
\newblock Ecological specialization and adaptive decay in digital organisms.
\newblock The American Naturalist 169:E1--E20.

\bibitem[{Pennell et~al.(2014)Pennell, Harmon, and Uyeda}]{Pennelletal2014}
Pennell, M.~W., L.~J. Harmon, and J.~C. Uyeda. 2014.
\newblock {Is there room for punctuated equilibrium in macroevolution?}
\newblock Trends in Ecology {\&} Evolution 29:23--32.

\bibitem[{Petersen(1975)}]{Petersen1975}
Petersen, R. 1975.
\newblock Paradox of plankton - equilibrium hypothesis.
\newblock The American Naturalist 109:35--49.

\bibitem[{Plucain et~al.(2014)Plucain, Hindre, Le~Gac, Tenaillon, Cruveiller,
  Medigue, Leiby, Harcombe, Marx, Lenski, and Schneider}]{Plucainetal2014}
Plucain, J., T.~Hindre, M.~Le~Gac, O.~Tenaillon, S.~Cruveiller, C.~Medigue,
  N.~Leiby, W.~R. Harcombe, C.~J. Marx, R.~E. Lenski, and D.~Schneider. 2014.
\newblock {Epistasis and Allele Specificity in the Emergence of a Stable
  Polymorphism in Escherichia coli}.
\newblock Science .

\bibitem[{Rainey and Travisano(1998)}]{RaineyTravisano1998}
Rainey, P.~B., and M.~Travisano. 1998.
\newblock Adaptive radiation in a heterogeneous environment.
\newblock Nature 394:69--72.

\bibitem[{Roy and Chattopadhyay(2007)}]{RoyChattopadhyay2007}
Roy, S., and J.~Chattopadhyay. 2007.
\newblock Towards a resolution of 'the paradox of the plankton': A brief
  overview of the proposed mechanisms.
\newblock Ecological Complexity 4:26--33.

\bibitem[{Schippers et~al.(2001)Schippers, Verschoor, Vos, and
  Mooij}]{Schippersetal1974}
Schippers, P., A.~Verschoor, M.~Vos, and W.~Mooij. 2001.
\newblock {Does "supersaturated coexistence" resolve the "paradox of the
  plankton"?}
\newblock Ecology Letters 4:404--407.

\bibitem[{Shoresh et~al.(2008)Shoresh, Hegreness, and Kishony}]{Shoresh2008}
Shoresh, N., M.~Hegreness, and R.~Kishony. 2008.
\newblock Evolution exacerbates the paradox of the plankton.
\newblock Proceedings of the National Academy of Sciences of the United States
  of America 105:12365--12369.

\bibitem[{Sommer(1984)}]{Sommer1984}
Sommer, U. 1984.
\newblock The paradox of the plankton: Fluctuations of phosphorous availability
  maintain diversity of phytoplankton in flow-through cultures.
\newblock Limnology and Oceanography 29:633--636.

\bibitem[{Tilman(1982)}]{Tilman1982}
Tilman, D. 1982.
\newblock Resource Competition and Community Structure.
\newblock Princeton University Press.

\bibitem[{{Van Valen}(1976)}]{VanValen1976}
{Van Valen}, L. 1976.
\newblock Ecological species, multispecies, and oaks.
\newblock Taxon 25:233--239.

\bibitem[{Weissing et~al.(2011)Weissing, Edelaar, and van
  Dorn}]{Weissingetal2011}
Weissing, F.~J., P.~Edelaar, and G.~S. van Dorn. 2011.
\newblock {Adaptive speciation theory: a conceptual review}.
\newblock Behavioral Ecology and Sociobiology 65:461.

\bibitem[{Wilkins(2007)}]{Wilkins2007}
Wilkins, J.~S. 2007.
\newblock The concept and causes of microbial species.
\newblock Studies in History and Philosophy of the Life Sciences 28:389--408.

\end{thebibliography}

\newpage

\begin{table}[htp]
{\begin{tabular}{ c  r  l  c  c  c  c }
\hline
log$\lambda$ & $\sigma_1$ & $\sigma_2$ & $\langle n_t \rangle^a$ & $B^a$ & $\langle n_t \rangle^b$ & $B^b$\\ \hline
-1 & ~1 & 0.1 & $1.35 \pm 0.49$ & 1.00 & $2.35 \pm 0.49$ & 0.45\\
-2 & ~1 & 0.1 & $2.05 \pm 0.76$ & 0.99 & $4.20 \pm 0.83$ & 0.37\\
-3 & ~1 & 0.1 & $2.25 \pm 0.85$ & 0.98 & $4.25 \pm 0.72$ & 0.36\\
-4 & ~1 & 0.1 & $1.95 \pm 0.89$ & 0.99 & $4.40 \pm 0.50$ & 0.36\\
-1 & 10 &  1 & $1.05 \pm 0.22$ & 0.81 & $1.20 \pm 0.41$ & 0.11\\
-2 & 10 &  1 & $1.95 \pm 0.76$ & 0.91 & $7.25 \pm 0.64$ & 0.13\\
-3 & 10 &  1 & $2.50 \pm 0.83$ & 0.91 & $8.85 \pm 0.37$ & 0.12\\
-4 & 10 &  1 & $2.65 \pm 0.81$ & 0.89 & $8.90 \pm 0.31$ & 0.12\\ \hline
\end{tabular}}
\caption{The effect of trait-lethal mutations is very strong. The degree of specialization is much lower and niche breadth is much higher when mutation can only increase or decrease resource affinity by $0.1$. $\mu=0.05$. $^a$ Without trait-lethals. $^b$ With trait-lethals.}
\label{table_nolethals}
\end{table}

\begin{table}[htp]
{ \begin{tabular}{ c  l  l  c  c  l }  \hline
log$\lambda$ & $\sigma_1$ & $\sigma_2$ & $\langle n_t \rangle^a$ & $\langle n_t \rangle^b$ & \multicolumn{1}{c}{$\langle n_t \rangle^c$}\\ \hline
-2 & 10  & 1    & $8.90 \pm 0.31$ & $7.25 \pm 0.64$ & $1.00 \pm 0.00$\\
-3 & ~0.1 & 0.01 & $3.25 \pm 1.16$ & $2.75 \pm 0.64$ & $1.60 \pm 0.50$\\
-3 & ~1   & 0.1  & $5.25 \pm 1.21$ & $4.25 \pm 0.72$ & $1.60 \pm 0.50$\\
-3 & 10  & 1    & $8.90 \pm 0.31$ & $8.85 \pm 0.37$ & $1.40 \pm 0.50$\\ \hline
\end{tabular}}
\caption{The population needs to be of a certain size in order for specialists to evolve. A population size of $N=100$ is too small to accommodate different ecotypes, while$1000$ is large enough that selection can distinguish between different phenotypes and negative frequency-dependent selection is able to sustain diversity. Increasing $N$ beyond $1000$ has only a small positive effect on the degree of specialization.  $\mu=0.05$. $^a$ $N=5000$. $^b$ $N=1000$. $^c$ $N=100$.}
\label{table_N}
\end{table}

\begin{table}[htp]
{\begin{tabular}{ c  l  l  c  c  c }  \hline
log$\lambda$ & $\mu$ & $\sigma_1$ & $\sigma_2$ & $\langle n_t \rangle^a$ & $\langle n_t \rangle^b$\\ \hline
-3 & 0.05  &  1 &      0.1 &  $4.30    \pm   0.80$ &  $4.25 \pm 0.72$\\
-3 & 0.01  &  1 &      0.1 &  $3.00    \pm   0.65$ &  $3.00 \pm 0.73$\\ \hline
\end{tabular}}
\caption{A comparison between simulations where the initial homogeneous population consists of specialists or generalists. The number of ecotypes is not different between the two initial conditions. $^a$ starting with generalists. $^b$ starting with specialists.}
\label{table_generalists}
\end{table}

\begin{table}[htp]
{\begin{tabular}{ l l l l l l l l l l }
\hline
blue & 0 &	0 &	0.3 &	0 &	0.4 &	0 &	0 &	0 &	0.2\\
red & 0 &	0.3 &	0.3 &	0 &	0.3 &	0 &	0.1 &	0 &	0.1\\
green & 0 &	0.1 &	0 &	0.4 &	0 &	0 &	0.4 &	0 &	0\\
cyan & 0.2 &	0.1 &	0 &	0 &	0 &	0.4 &	0 &	0.4 &	0.1\\ \hline
\end{tabular}}
\caption{Four stable phenotypes after 5,000 updates. Colors refer to the lineages in figure~\ref{fig_four_LODs}.}
\label{table_four_phenotypes}
\end{table}

\begin{table}[htp]
{\tiny \begin{tabular}{r | l l l l l l l l l | l }
{\normalsize update} & \multicolumn{9}{c}{{\normalsize phenotype}} & \multicolumn{1}{c}{\normalsize s}\\ \hline
~ & \multicolumn{9}{l}{{\normalsize \textcolor[rgb]{0.00,0.00,1.00}{blue lineage}}}  & ~ \\
1 & 0.1 & 0 & 0 & 0 & 0 & 0 & 0 & 0 & 0 &  \\
15 & 0.1 & 0 & 0 & 0 & 0 & 0 & 0.1 & 0 & 0.1 & 11.271\\
40 & 0.1 & 0 & 0.1 & 0 & 0 & 0 & 0.1 & 0 & 0.1 &  ~0.30965\\
89 & 0.1 & 0 & 0.1 & 0 & 0.1 & 0 & 0 & 0 & 0.1 &  ~0.039084\\
237 & 0 & 0 & 0.1 & 0 & 0.1 & 0 & 0 & 0 & 0.1 &  ~0.019431\\
612 & 0 & 0 & 0.2 & 0 & 0.1 & 0 & 0 & 0 & 0.1 &  ~0.29279\\
1124 & 0 & 0.1 & 0.2 & 0 & 0.2 & 0 & 0 & 0 & 0.1 &  ~0.27836\\
1393 & 0 & 0.2 & 0.2 & 0 & 0.2 & 0 & 0 & 0 & 0.1 &  ~0.10041\\
2121 & 0 & 0.2 & 0.2 & 0 & 0.3 & 0 & 0 & 0 & 0.1 &  ~0.10848\\
2658 & 0 & 0.2 & 0.2 & 0 & 0.3 & 0 & 0 & 0 & 0.2 &  ~0.050309\\
2911 & 0 & 0.2 & 0.3 & 0 & 0.3 & 0 & 0 & 0 & 0.2 & ~0.077457\\
2916 & 0 & 0.1 & 0.3 & 0 & 0.3 & 0 & 0 & 0 & 0.2 & -0.052489\\
2999 & 0 & 0.1 & 0.3 & 0.1 & 0.3 & 0 & 0 & 0 & 0.2 & ~0.030072\\
3196 & 0 & 0 & 0.3 & 0 & 0.3 & 0 & 0 & 0 & 0.2 & -0.044106\\
3249 & 0 & 0 & 0.3 & 0.1 & 0.3 & 0 & 0 & 0 & 0.2 & ~0.037581\\
3541 & 0 & 0 & 0.3 & 0.1 & 0.4 & 0 & 0 & 0 & 0.2 & ~0.067862\\
3802 & 0 & 0 & 0.3 & 0.2 & 0.4 & 0 & 0 & 0 & 0.2 & ~0.06681\\
4243 & 0 & 0 & 0.3 & 0.1 & 0.4 & 0 & 0 & 0 & 0.2 & -0.067928\\
4878 & 0 & 0 & 0.3 & 0 & 0.4 & 0 & 0 & 0 & 0.2 & -0.027721\\
 & \multicolumn{9}{l}{{\normalsize \textcolor[rgb]{0.98,0.00,0.00}{red lineage}}}  &  \\
1	& 0.1 & 0 & 0 & 0 & 0 & 0 & 0 & 0 & 0 & \\
15	& 0.1 & 0 & 0 & 0 & 0 & 0 & 0.1 & 0 & 0.1 & 11.271\\
40	& 0.1 & 0 & 0.1 & 0 & 0 & 0 & 0.1 & 0 & 0.1 & ~0.30965\\
89	& 0.1 & 0 & 0.1 & 0 & 0.1 & 0 & 0 & 0 & 0.1 & ~0.039084\\
237	& 0 & 0 & 0.1 & 0 & 0.1 & 0 & 0 & 0 & 0.1 & ~0.019431\\
612	& 0 & 0 & 0.2 & 0 & 0.1 & 0 & 0 & 0 & 0.1 & ~0.29279\\
1124	& 0 & 0.1 & 0.2 & 0 & 0.2 & 0 & 0 & 0 & 0.1 & ~0.27836\\
1393	& 0 & 0.2 & 0.2 & 0 & 0.2 & 0 & 0 & 0 & 0.1 & ~0.10041\\
2121	& 0 & 0.2 & 0.2 & 0 & 0.3 & 0 & 0 & 0 & 0.1 & ~0.10848\\
2658	& 0 & 0.2 & 0.2 & 0 & 0.3 & 0 & 0 & 0 & 0.2 & ~0.050309\\
2911	& 0 & 0.2 & 0.3 & 0 & 0.3 & 0 & 0 & 0 & 0.2 & ~0.077457\\
3028	& 0 & 0.3 & 0.3 & 0 & 0.3 & 0 & 0 & 0 & 0.2 & ~0.046359\\
3095	& 0 & 0.3 & 0.3 & 0 & 0.3 & 0 & 0 & 0 & 0 & -0.040099\\
3684	& 0 & 0.3 & 0.3 & 0 & 0.3 & 0 & 0.1 & 0 & 0 & ~0.036444\\
4850	& 0 & 0.3 & 0.3 & 0 & 0.3 & 0 & 0.1 & 0 & 0.1 & -0.0017963\\
 & \multicolumn{9}{l}{{\normalsize \textcolor[rgb]{0.00,0.59,0.00}{green lineage}}}  &  \\
1	& 0.1 & 0 & 0 & 0 & 0 & 0 & 0 & 0	& 0 &	\\
15	& 0.1 & 0 & 0 & 0 & 0 & 0 & 0.1 & 0 & 0.1 & 11.271\\
67	& 0.1 & 0.1 & 0 & 0 & 0 & 0 & 0.1 & 0 & 0.1 & ~0.37720\\
75	& 0.1 & 0.1 & 0 & 0 & 0 & 0 & 0.1 & 0.1 & 0.1 & ~0.18758\\
225	& 0.1 & 0.1 & 0 & 0.1 & 0 & 0 & 0.1 & 0.1 & 0.1 & ~0.17634\\
329	& 0.1 & 0.1 & 0 & 0.1 & 0 & 0 & 0.2 & 0.1 & 0.1 & ~0.08945\\
548	& 0.1 & 0.1 & 0 & 0.1 & 0 & 0.1 & 0.2 & 0.1 & 0.1 & ~0.16142\\
642	& 0 & 0.1 & 0 & 0.1 & 0 & 0.1 & 0.2 & 0.1 & 0.1 & -0.0049659\\
800	& 0 & 0.1 & 0.1 & 0.1 & 0 & 0.1 & 0.2 & 0.1 & 0.1 & ~0.094182\\
1060	& 0 & 0.1 & 0 & 0.1 & 0 & 0.1 & 0.2 & 0.1 & 0.1 & -0.087319\\
1230	& 0 & 0.1 & 0 & 0.1 & 0 & 0.1 & 0.3 & 0.1 & 0.1 & ~0.076287\\
1741	& 0 & 0.1 & 0 & 0.2 & 0 & 0.1 & 0.3 & 0.1 & 0.1 & ~0.10556\\
2704	& 0 & 0.1 & 0 & 0.2 & 0 & 0.1 & 0.4 & 0.1 & 0.1 & ~0.05857\\
2725	& 0 & 0.1 & 0 & 0.2 & 0 & 0 & 0.4 & 0.1 & 0.1 & -0.025019\\
2855	& 0 & 0.1 & 0 & 0.2 & 0 & 0 & 0.4 & 0.1 & 0.2 & ~0.052496\\
2995	& 0 & 0.1 & 0 & 0.2 & 0 & 0 & 0.4 & 0.1 & 0 & -0.058368\\
3570	& 0 & 0.1 & 0 & 0.3 & 0 & 0 & 0.4 & 0.1 & 0 & ~0.087326\\
4859	& 0 & 0.1 & 0 & 0.4 & 0 & 0 & 0.4 & 0 & 0 & ~0.068163\\
 & \multicolumn{9}{l}{{\normalsize \textcolor[rgb]{0.00,0.84,0.84}{cyan lineage}}}  &  \\
1 & 0.1 & 0 & 0 & 0 & 0 & 0 & 0 & 0	& 0 & \\
15 & 0.1 & 0 & 0 & 0 & 0 & 0 & 0.1 & 0 & 0.1 & 11.271\\
67 & 0.1 & 0.1 & 0 & 0 & 0 & 0 & 0.1 & 0 & 0.1 & ~0.3772\\
75 & 0.1 & 0.1 & 0 & 0 & 0 & 0 & 0.1 & 0.1 & 0.1 & ~0.18758\\
134 & 0.1 & 0.1 & 0 & 0 & 0 & 0 & 0.1 & 0.2 & 0.1 & ~0.20596\\
241 & 0.1 & 0.1 & 0 & 0 & 0.1 & 0 & 0.1 & 0.2 & 0.1 & ~0.13729\\
369 & 0.1 & 0.1 & 0 & 0 & 0.1 & 0.1 & 0.1 & 0.2 & 0.1 & ~0.19802\\
377 & 0.1 & 0.1 & 0 & 0 & 0.1 & 0.2 & 0.1 & 0.2 & 0.1 & ~0.17998\\
655 & 0.1 & 0 & 0 & 0 & 0.1 & 0.2 & 0.1 & 0.2 & 0.1 & -0.042485\\
1776 & 0.1 & 0 & 0 & 0 & 0.1 & 0.2 & 0.1 & 0.3 & 0.1 & ~0.0667\\
1868 & 0.1 & 0 & 0 & 0 & 0.1 & 0.2 & 0.1 & 0.4 & 0.1 & ~0.055831\\
2112 & 0.2 & 0 & 0 & 0 & 0.1 & 0.2 & 0.1 & 0.4 & 0.1 & ~0.076066\\
2115 & 0.2 & 0 & 0 & 0 & 0.1 & 0.3 & 0.1 & 0.4 & 0.1 & ~0.04995\\
2283 & 0.2 & 0 & 0 & 0 & 0.1 & 0.4 & 0.1 & 0.4 & 0.1 & ~0.039907\\
2669 & 0.2 & 0 & 0 & 0 & 0.1 & 0.4 & 0 & 0.4 & 0.1 & -0.00022735\\
4124 & 0.2 & 0 & 0 & 0 & 0 & 0.4 & 0 & 0.4 & 0.1 & -0.0067946\\
4545 & 0.2 & 0.1 & 0 & 0 & 0 & 0.4 & 0 & 0.4 & 0.1 & ~0.013724\\ \hline
\end{tabular}}
\caption{Complete mutational history of all four lines of descent in table~\ref{table_four_phenotypes}. The selection coefficient is $s=W'/W-1$, where $W'$ is the fitness of the offspring, and $W$ is the fitness of the parent. 
}
\label{table_four_lod}
\end{table}

\begin{table}[htp]
{\begin{tabular}{ l  c  l  l  l  c  l  l}
\hline
log$\lambda$ & $\mu$ & $\sigma_1$ & $\sigma_2$ & \multicolumn{1}{c}{$\langle n_t \rangle$}  & $B$ & \multicolumn{1}{c}{$N$}\\ \hline
-3 & 0.01 & ~1 & 0.1 & $2.90 \pm 0.55$ & $0.62$ & ~$720$ constant \\
-3 & 0.01 & ~1 & 0.1 & $2.95 \pm 0.51$ & $0.63$ & ~$721$ variable \\
-3 & 0.05 & ~1 & 0.1 & $4.45 \pm 0.76$ & $0.40$ & ~$750$ constant\\
-3 & 0.05 & ~1 & 0.1 & $4.20 \pm 0.52$ & $0.40$ & ~$755$ variable \\
-2.4 & 0.05 & 10 & 1 & $9.00 \pm 0.00$ & $0.11$ & $2520$ constant \\
-2.4 & 0.05 & 10 & 1 & $8.75 \pm 0.55$ & $0.12$ & $2520$  variable\\
\hline
\end{tabular}}
\caption{Allowing the population to vary in size does not affect the degree of specialization. The number of ecotypes stays the same between constant and variable population size simulations for each parameter set. 
The population size given is the mean of twenty runs, which varies very little; values range between 695 and 799 for the mean of 755, 696 and 784 for the mean of 721, and from 2469 to 2613 for the mean of 2520.}
\label{table_varpop}
\end{table}

\end{document}